 \documentclass[12pt,preprint]{aastex}

\newcommand{\sst}{{\it Spitzer Space Telescope}}

\newcommand{\kms}{km~s$^{-1}$}

\shorttitle{IR Galaxy Morphologies}
\shortauthors{Pahre et al.}

\begin{document}

\title{Mid-Infrared Galaxy Morphology Along the Hubble Sequence}

\author{Michael A. Pahre,
  M. L. N. Ashby,
  G. G. Fazio, and
  S. P. Willner}

\affil{Harvard-Smithsonian Center for Astrophysics, 60 Garden Street, Cambridge, MA  
  02138\email{mpahre, mashby, gfazio, swillner@cfa.harvard.edu} }

\begin{abstract}
The mid-infrared emission 
from 18 nearby galaxies imaged with the IRAC instrument on \sst\
samples the spatial distributions of the reddening-free 
stellar photospheric emission and the warm dust in the ISM.
These two components provide a new framework for galaxy
morphological classification, in which the presence of spiral arms
and their emission strength relative to the starlight can be measured
directly and with high contrast.
Four mid-infrared classification methods are explored, three of which
are based on quantitative global parameters (colors, bulge-to-disk
ratio) similar to those used in the past for optical studies;
in this limited sample, all correlate well with traditional 
$B$-band classification.
We suggest reasons why infrared classification may
be superior to optical classification.
\end{abstract}


\section{Introduction}
The morphological classification scheme introduced by Hubble (1926,
1936), based on blue-light images, 
has been modified periodically over the years (e.g., Sandage
1961; de Vaucouleurs 1959; Kormendy 1979; Buta 1995) 
but remains a fundamental method by which
astronomers continue to sort and compare galaxies.  
From a contemporary viewpoint, the scheme is useful because it closely follows the
fundamental distinction between galaxies with and without significant 
amounts of interstellar matter.  
Within each major class (spirals and irregulars on the one hand,
ellipticals and lenticulars on the other), the subclass
reflects the history of each galaxy, although the details are still a
topic of active study.  
Furthermore, within the spiral-irregular branch,
subclass is closely correlated with the number and luminosity of young stars
relative to old stars.


Hubble morphological classification is essentially a visual exercise
which is sometimes assigned a numerical index (such as the $T$ index in
the RC3; de Vaucouleurs et al. 1991) in order to facilitate comparisons
with various quantitative global parameters.
Despite the essentially qualitative process, there is general
agreement among many human classifiers \cite{naim95}.


Different classification schemes have been developed over the years,
such as spectral \cite[absorption versus emission line;][]{morgan57,bromley98},
ultraviolet \cite{marcum01}, and near-infrared \cite{jarrett00,block99}.
Detailed comparisons have also been made between the optical classifications
and various global galaxy parameters, such as:
color \citep[blue versus red;][]{buta94,devauc60};
light distribution via the bulge-to-disk-ratio \cite{vdb76},
concentration index (Okamura et al. 1984; Abraham et al. 1994), and
galaxy asymmetry \cite{abraham96};
and trained neural network schemes \cite{odewahn96}.
These various methods often show good correspondence among them
\cite{roberts94}, but there are significant and systematic differences
among at least some (Connolly et al. 1995; Kochanek, Pahre, \& Falco 2000).
So far, no alternate scheme appears to be as directly tied to the
composition and history of a galaxy as Hubble's.

In this paper, we revisit the morphological classification scheme using
images with resolution and sensitivities similar to those used for
traditional optical classifications---but at mid-infrared wavelengths.
While some nearby galaxies could be resolved by previous infrared space
missions operating at these wavelengths, and equal or better spatial 
resolution can be obtained from the ground,
the combined wide-field coverage and sensitive infrared imaging detectors
of the IRAC instrument on \sst\ permits us to explore an entirely new 
region of parameter space for imaging nearby galaxies.
These images demonstrate a new approach to galaxy classification based on
the properties of the galaxy \emph{interstellar medium} relative to its starlight.

\section{Sample}

The 16 galaxies presented here are a small but
representative subset of a sample of about 100 scheduled to be
observed.  These galaxies were drawn from a complete sample from 
Ho, Filippenko, \& Sargent (1997), 
which contains nearly 500 galaxies in the northern sky at $B_T < 12.5$~mag.
A subsample of approximately 100 galaxies was drawn from it under the constraints:
(1) the galaxies should have small sizes (less than $7$~arcmin), so that they
would fit in the instantaneous instrument field-of-view; and
(2) the two-parameter space of luminosity and morphological type is fully
spanned.
The galaxies were randomly drawn within these constraints.
Several low surface brightness galaxies were added to the subsample, as the parent sample
was judged to be missing such galaxies.  
The sample will be more fully described in a future paper.
It is fortunate that the sample to date nearly fully spans the classical
morphological sequence (Sa and Sab are missing).
Two additional galaxies (NGC~3031 and NGC~300) were observed for
Early Release Observations; the first one is in the parent sample, while the
latter is southern and is not.
Both observations are described elsewhere in this issue (Willner et al. 2004;
Helou et al. 2004), and are only included in the color and bulge-to-disk-ratio
plots.
We adopt the RC3 optical classifications \cite{rc3} throughout this paper for comparison
purposes, with the exception of NGC~5363:  
it is mis-classified as I0?, but is most likely intended to be the type S0pec.  
We have adopted E/S0pec for it.

\section{Observations}


The data were taken with the InfraRed Array Camera 
\cite[IRAC;][]{irac} on the {\it Spitzer Space Telescope} \cite{sstpaper} 
during the first four IRAC campaigns of normal operations (2003 December -- 2004
March).  
\dataset[ads/sa.spitzer#0004431104]{Sixteen galaxies}
were imaged, each using five exposures of
either 12 or 30 seconds each on the source in all four detectors.
The standard pipeline BCD (Basic Calibrated Data, version 8.9--9.5)
products were used in the reductions.
The first 5.8~\micron\ image had an additional
``first-frame delta skydark'' subtracted from it, which was based on
an independent dataset.
The bias level for the 5.8~\micron\ images was
significantly variable such that the median value of the background plus bias 
(which was iteratively determined via a sigma-clipping algorithm) had to be
subtracted from each individual frame before mosaicing.

A mosaic of each target in each bandpass was made using custom
software operating under the {\sc iraf} environment and using the
{\sc wcstools} package.  Optical distortion was removed in the
process, and the images were subpixellated to $0.86$~arcsec, which is
a linear reduction by a factor of $\sqrt{2}$ in each dimension.  A
first iteration mosaic was constructed using a median combination of
the individual frames.  Each separate BCD frame was then compared
with the mosaic, and differences much larger than the amount expected from
the known noise were assumed to be cosmic rays or other artifacts.
Pixels affected were flagged as invalid, and the resulting
cleaned frames were then combined into the final mosaic. The
resulting 3.6 and 8~\micron\ images are shown in Figure~1.  The
4.5 and 5.8~\micron\ images look similar to 3.6 and 8.0~\micron,
respectively. 

\placefigure{\fig1}

Quantitative data in this paper are given in instrumental magnitudes
relative to Vega.  The existing calibration of IRAC data is based on
point source observations \cite{irac}.   
Extended sources will have
aperture corrections different from those for point sources, and the
correct values are still unknown.  
In particular, for the elliptical
galaxy NGC~777, which shows the least evidence for any dust emission,
we assume a pure stellar spectrum matching that of an M0~III star.
This stellar class is chosen in the mid-infrared, since the $K$-band
light of old stellar populations has been shown to be dominated by the 
M~giants \cite{frogel78}.
The resulting corrections, representing the ratio of point source
flux in a 12.2~arcsec aperture to that in an infinite aperture, are
$(0.09,0.05,0.17,0.31)$~mag.  These corrections agree within
$0.03$~mag with preliminary estimates based on the flux at $r >
12.2$~arcsec for observations of Fomalhaut, except for
5.8~\micron, where the galaxy magnitudes implied by the Fomalhaut
data are $\sim 0.15$ mag fainter.  
If NGC~777 does have dust emission after all, then the other galaxies will have
larger emission than derived here.


\section{Analysis}

For each galaxy, the $[3.6]-[4.5]$ color is nearly constant with
radius and is consistent with stellar photosphere emission. Dust
emission is likely to be important at longer wavelengths.
The stellar emission was subtracted from the $5.8$ and $8.0~\mu$m
images using the $3.6$ and $4.5~\mu$m images suitably scaled to match
the theoretical colors of M0~III stars: $[3.6]-[4.5]=-0.15,
[4.5]-[5.8] = +0.11, [5.8]-[8.0] = +0.04$~mag (M. Cohen and
T. Megeath, private comm.).  The resultant images are referred to as
``non-stellar emission'' and are shown in Figure~1.

Surface photometry was measured on the galaxies using the IRAF task
{\sc ELLIPSE}, following the methodology of Pahre
(1999).  
The full fitting was first done on the $3.6
\mu$m image, then the flux at each isophote was measured for the
other wavelengths using the isophotal shapes defined from the $3.6
\mu$m image.  
Circular aperture magnitudes
were also measured for each galaxy.  Corrections for flux lost due to
the PSF were applied to the surface brightness and aperture magnitude profiles
in the manner described in Pahre (1999).
One-dimensional models of $r^{1/4}$ bulge, exponential disk, and the
two combined were fit to the isophotal surface brightness profiles
and the circular aperture magnitude curves of growth.  
The $B/D$ values from the two approaches were averaged, and half the difference
is taken as the uncertainty.

The morphological types, photometric parameters, and model fits for each model
are tabulated in Table~\ref{tab1}.
The near-IR and mid-IR colors in the table were measured within matched apertures
defined by the $\mu_K = 20$~mag~arcsec$^{-2}$ isophote from the 2MASS
XSC.  
Uncertainties are tabulated for the model fits to $B/D$; uncertainties for
the true colors are currently heavily dominated by the systematic effects both
of the stellar calibration and the extended source calibration based on
NGC~777 adopted here.

\placetable{\ref{tab1}}


\section{Infrared Galaxy Classification}


\subsection{Morphology of the $\lambda = 8.0 \mu$m Emission}


The classification is based simply on the visual appearance
at 8~\micron\ or perhaps even better, on the visual appearance of
non-stellar emission at 8~\micron, as shown in Figure~1.  The reason
is easy to see: the dust emission tracks interstellar matter, which
is predominantly resident in spiral arms.  Arm/interarm contrast is
significantly higher in 8~\micron\ images than in visible images.
While not directly shown here, we note that ultraviolet light 
also shows high contrast, but can be affected significantly by
dust obscuration.


\subsection{Infrared Colors of the Stellar Population}

The IRAC color $[3.6]-[4.5]$ varies systematically with galaxy
morphological type as shown in Figure~2.  This is probably a
population age effect: young A0V stars have color of zero, an
intermediate-age population matches the \emph{bluer} color of a K0~III
star (-0.06~mag), and an old stellar population matches the yet
\emph{bluer} colors for later-type K2~III-M0~III stars of $-0.14$ to
$-0.15$~mag.  
While this is a rather simplistic exploration of the intrinsic colors of the
stellar population, it does highlight a general feature of the $[3.6]-[4.5]$ color:
that it becomes \emph{redder} for galaxies of increasingly late types.

\placefigure{\fig2}

The amount by which the $[3.6]-[4.5]$ color changes with galaxy type $T$ is 
$\Delta([3.6]-[4.5])/\Delta T \sim 0.01$~mag. 
This is similar to the slope for the optical color $(V-R)$, 
but significantly smaller than other optical colors $(U-B)$, $(B-V)$,
and $(V-I)$ \cite{buta95b}.
While this might imply that IR classification based on $[3.6]-[4.5]$ is more
prone to the effects of photometric errors, the optical colors are affected
by the systematic error of internal reddening and hence by inclination effects.
For this reason, we argue that IR colors are more representative of the 
properties of the underlying stellar populations.

The near-infrared color $K_s - [3.6]$ shows significantly more scatter than
the IRAC-only colors.
Part of this is due to the larger photometric errors from the 2MASS measurements,
which range from 0.02 to 0.11~mag.
Another part of the large scatter may lie in imperfect matching of the isophotal
aperture between 2MASS and IRAC.

\subsection{Infrared Colors of the Non-Stellar Emission}

The IRAC color $[3.6]-[8.0]$ varies with galaxy type even more
strongly than $[3.6]-[4.5]$ (Figure~2).  This is because the $8.0
\mu$m filter is detecting the warm dust, probably emitting mostly via
the PAH emission features at $\lambda = 7.7$ and $8.6 \mu$m (Gillett, Forrest, \& Merrill  1973).
Galaxies of increasing morphological type $T$ show
increasing ratios of PAH emission at $8.0 \mu$m to stellar mass at
$3.6 \mu$m, hence this color becomes \emph{redder} with $T$.  The
fraction of 8~\micron\ light  from non-stellar emission ranges from
$<$10\% for the early-type galaxies to $> 90$\% for the late types.

\subsection{Infrared Bulge-to-Disk Ratios}

Since starlight at $\lambda = 3.6 \mu$m very nearly follows the
Rayleigh-Jeans limit of blackbody emission for $T > 2000$~K, the
colors of both early- and late-type stars are all nearly the same.
There is virtually no dust extinction at this wavelength either, as
any standard extinction law predicts only a few percent of the
extinction compared to optical wavelengths.  The $3.6 \mu$m light
therefore traces the stellar mass distribution free of dust
obscuration effects. 

The bulge-to-disk-ratios for the galaxies progressively decrease with
galaxy morphological type, demonstrating that
there is an increasing \emph{stellar} mass fraction in the disk
component of the later-type galaxies.  Unlike optical wavelengths,
this effect cannot be ascribed to the emission of a small number of
luminous young stars which have little total mass.
Since the $3.6 \mu$m light is virtually all stellar, the bulge-to-disk-ratios
we derive here will match the $K$-band.

\section{Discussion of Individual Objects}

Galaxy NGC~5548 has a powerful AGN of type Sy~1.5 in its nucleus, which
appears very red relative to the starlight (Figure~1).  It has been studied
extensively as part of the {\sc agnwatch} project \cite{peterson02}.
No correction for the nuclear emission was done, hence the galaxy is
plotted as an open symbol in Figure~2.

A number of other galaxies show much weaker evidence, via both spatially
resolved point source nuclei and reddening trends in $[3.6]-[4.5]$, for active nuclei
(see Willner et al. 2004 for the case of M~81).
Spatial and spectral separation of AGN will be the subject of a future contribution.

Lenticular early-type galaxies NGC~5363, 1023, and 4203 all show evidence of
spatially resolved dust emission at $8.0 \mu$m, and are discussed in a 
companion paper \cite{pahre2004b}.


Nearly edge-on spiral NGC~5746 is classified by RC3 as SAB(rs)b/sp?, which reflects
considerable uncertainty due to its prominent dust lane.
The IRAC images easily penetrate the obscuration and demonstrate 
that it is SB(r)ab, 
with the presence of a large inner ring and an apparent bar roughly 
perpendicular to the plane of the sky.

\section{Summary}

IRAC images offer powerful new tools for studying both the stellar
and interstellar components of nearby galaxies.  Both the relative
amounts of these components and their spatial distributions can be
used to classify galaxies.  Early indications from this limited
sample are that such classifications correlate well with the
traditional Hubble scheme.  Three of four classification schemes
suggested are strictly quantitative and therefore can be applied
without personal biases; the fourth could probably be made
quantitative through Fourier analysis techniques.

The principal advantages of a mid-infrared classification scheme are:
\begin{enumerate}
\item The mid-IR traces both interstellar matter and starlight without the
    effects of extinction. 
\item The mid-IR dust emission, particularly the PAH feature at
    $\lambda = 7.7~\mu$m, is a clear tracer of the presence of
    interstellar matter.  The emission shows high contrast against
    stellar emission at the same wavelength.
\item The colors of stellar photospheres in the mid-IR vary only a
     small amount with population age or mass function, and 
     hence the stellar emission is a direct tracer of stellar
     mass.  The bulge-to-disk-ratios measured at $3.6$ and $4.5 \mu$m 
     therefore sample the mass ratio 
     of the stellar content, not a mixture of stellar content and
     recent massive star formation activity.
\item The stellar photospheric component is straightforward to model
     based on the $3.6$ and $4.5 \mu$m images, and hence provides a
     simple framework for separating the stellar from the ISM emission
     at $5.8$ and $8.0 \mu$m.
\end{enumerate}

The mid-IR emission is a more natural tracer of the
underlying astrophysics that gives rise to the galaxy sequence,
namely the star formation rate per unit stellar mass.  
Future work may include a quantitative investigation of the 
properties of galactic potentials, bars, and spiral arms.
As Figure~1 demonstrates, such galaxy features can be imaged strikingly 
well at mid-infrared wavelengths.


\acknowledgments

M.A.P. acknowledges NASA/LTSA grant \# NAG5-10777.
The IRAC GTO program is supported by  JPL Contract \# 1256790.
The IRAC instrument was developed under JPL Contract \# 960541.
This work is based on observations made with the \sst, which is
operated by the Jet Propulsion Laboratory, California Institute of
Technology under NASA contract 1407. 


Facilities:  \facility{Spitzer(IRAC)}, \facility{2MASS}.

\clearpage

\begin{figure}
\epsscale{0.80}
\plotone{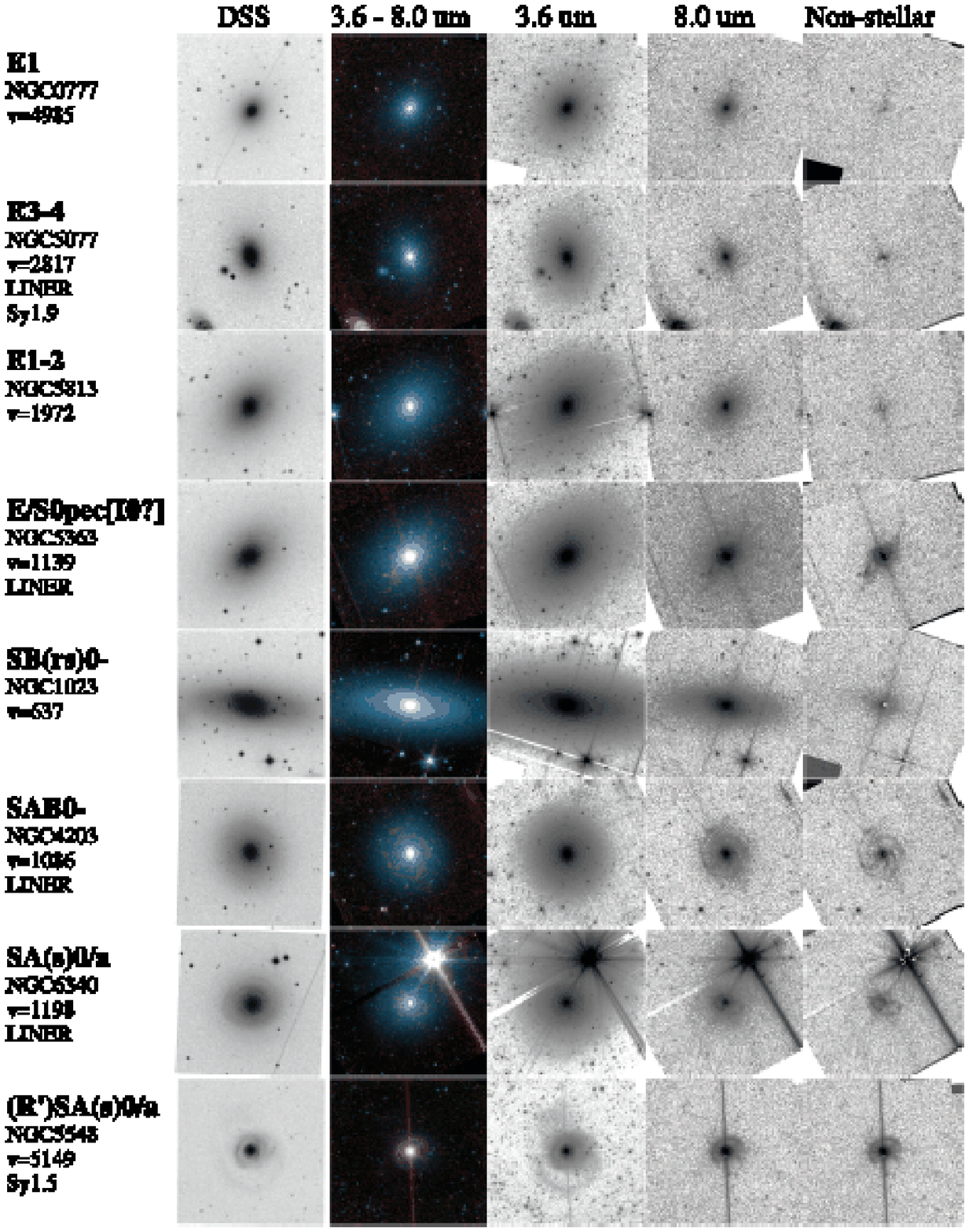}
\caption{
{\bf (PLATES 1 and 2.)}
Images of the galaxies sorted by increasing morphological type
$T$ (RC3).
Blue DSS images are provided for comparison.
Each panel is $5' \times 5'$ in size.
The IRAC color images are comprised of $3.6$ (``blue''),
$4.5$ (``green''), and $8.0$~\micron\ (``red'').
The $8.0 \mu$m ``non-stellar'' images have had the stellar light
subtracted from them (see text).
All IRAC images displayed in this figure use an identical 
square--root scaling and limits of 0--10~MJy/sr (greyscale) or
0--5~MJy/sr (color).
\label{fig1} }
\end{figure}

\clearpage
\plotone{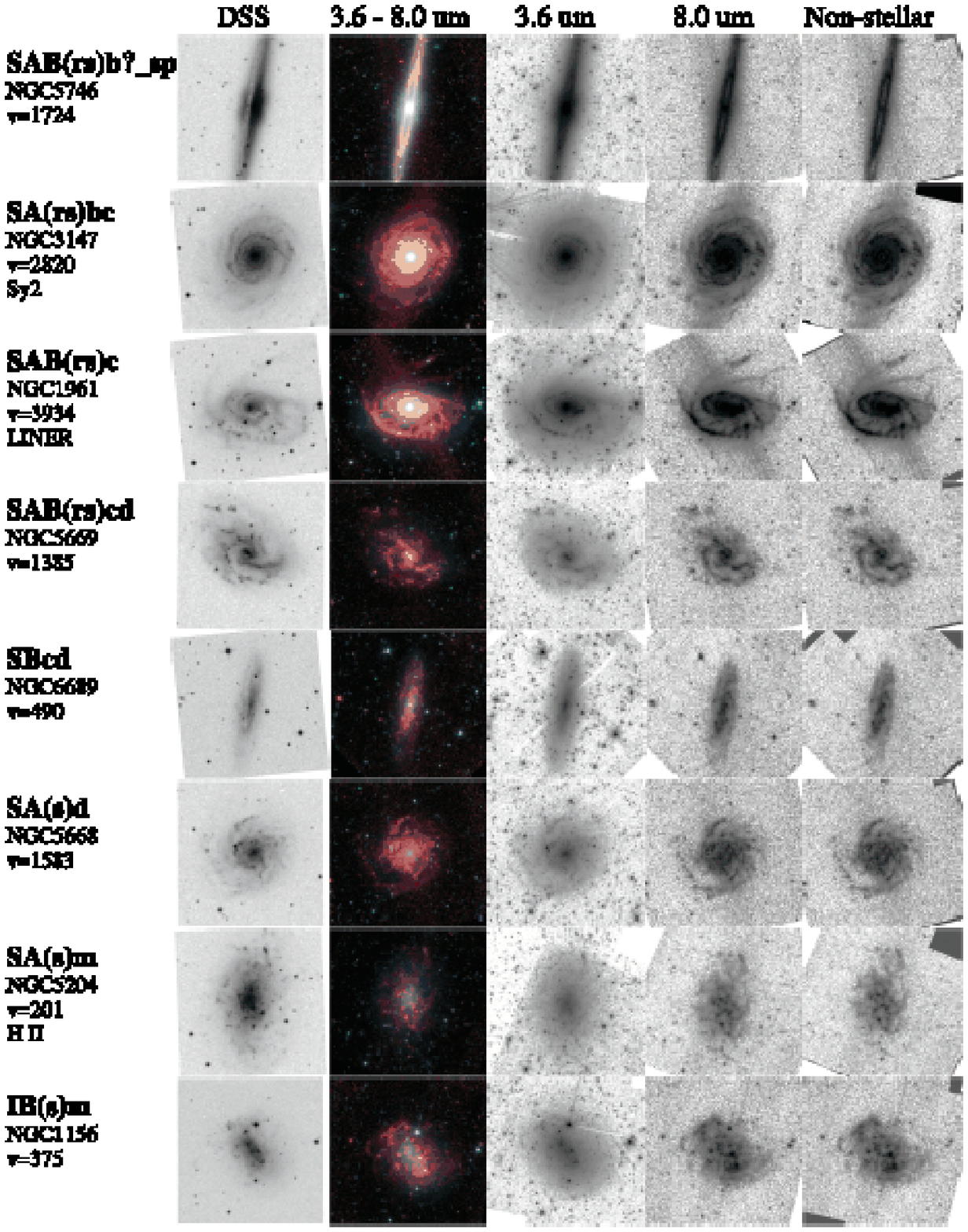}

\begin{figure}
\epsscale{0.75}
\plotone{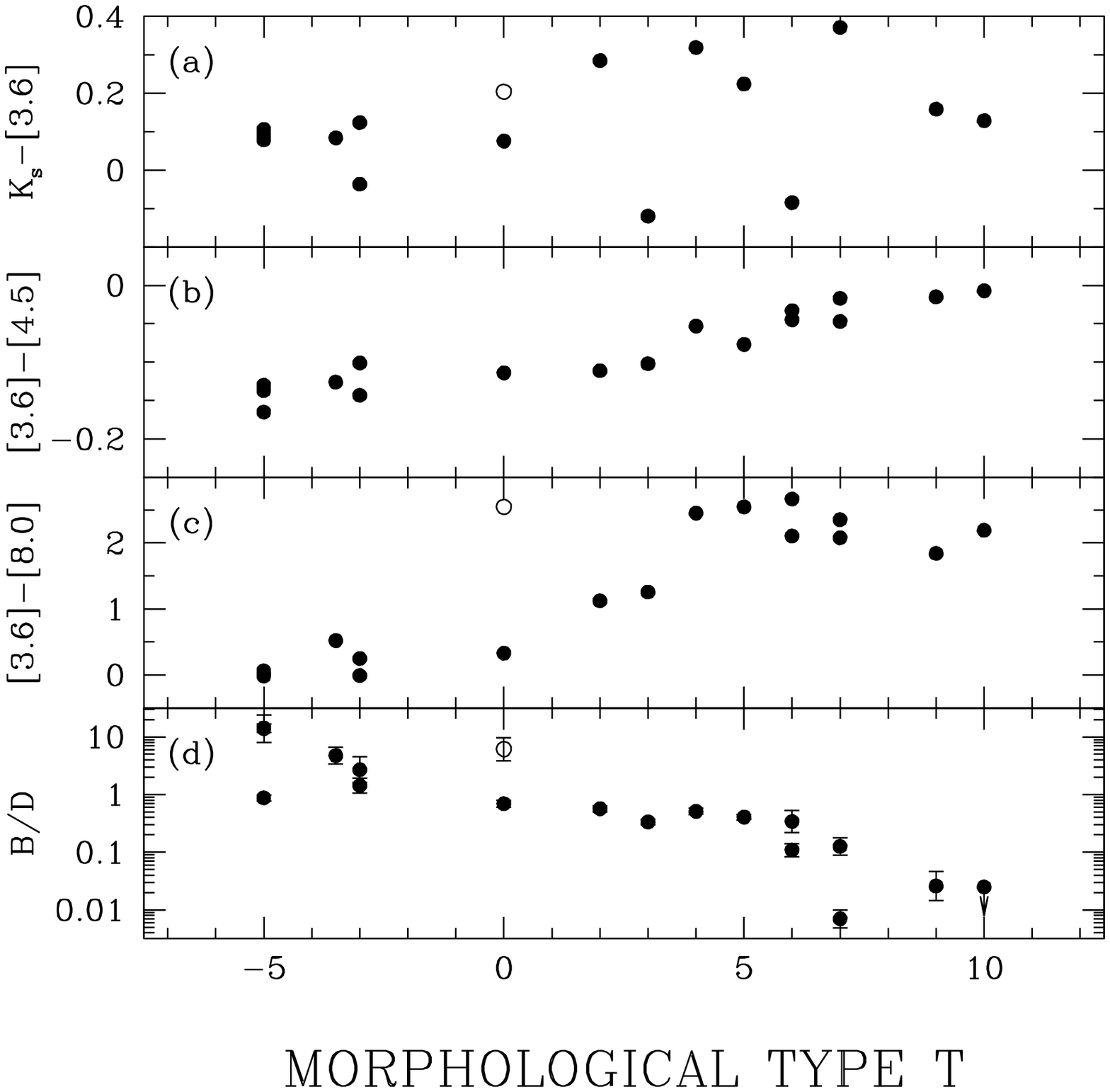}
\caption{Colors and bulge-to-disk ratios of the galaxies as a
function of morphological  type $T$.  (a) The 2MASS--IRAC color $K_s
- [3.6]$ evaluated at the $\mu_K = 20$~mag~arcsec$^{-2}$ isophote.
AGN-dominated NGC~5548 is plotted as an open symbol.
(b) The IRAC color $[3.6]-[4.5]$ evaluated at the same isophote.
Early-type galaxies are blue in this color and match the
colors expected for stars of type M0~III.  Late-type
galaxies are red in this color, matching the colors expected for
K0~III through A0~V stars.  The AGN-dominated NGC~5548 is off the plot
at $[3.6]-[4.5] = +0.55$~mag.
(c) Same for $[3.6]-[8.0]$.  The non-stellar
emission goes from near-zero contribution in
the early-type galaxies to the dominant emission for the
late-type galaxies.  ($[3.6]-[8.0] = 2.5$~mag implies
that the PAH emission is ten times greater than the stellar emission.)
(d) Bulge-to-disk ratio for the galaxies as measured in the IRAC $3.6
\mu$m images.  Emission in this bandpass is a direct tracer of
the stellar mass distribution, hence these bulge-to-disk ratios imply
similar ratios in the stellar masses in each galaxy component.  The
bulge-to-disk-ratio shows a strong decrease from early- to late-type
galaxies.  The plotted value for NGC~1156 ($T=10$, type Im) is an
upper limit; there is no evidence for a bulge component.
\label{fig2} }
\end{figure}

%

\clearpage

\begin{deluxetable}{llrrrrrrrrrr}
\tabletypesize{\scriptsize}
\rotate
\tablecaption{Galaxy Morphological, Photometric, and Model Fit Quantities  \label{tab1}}
\tablewidth{0pt}
\tablehead{
\colhead{Galaxy} & 
\colhead{Type} & 
\colhead{$T$} &
\colhead{$V_H$} & 
\colhead{$r_(K_{20})$} & 
\colhead{$b/a$} &
\colhead{$K_{20}$} &
\colhead{$K-[3.6]$} & 
\colhead{$[3.6]-[4.5]$} & 
\colhead{$[3.6]-[8.0]$} & 
\colhead{$B/D$} & \colhead{$\pm$} \\
\colhead{name} & \colhead{} & \colhead{} &
\colhead{\kms} & \colhead{arcsec} & \colhead{} & 
\colhead{mag} & \colhead{mag} &
\colhead{mag} & \colhead{mag} & 
\colhead{} 
}
\startdata

\dataset[ads/sa.spitzer#0004431104]{NGC~\phn777}  & E1 Sy2               &   -5.0 &   4985 &   51.1 & 0.840 &  8.508 &  0.092 & -0.165 &  0.004 &    14.320 &   2.634 \\ 
\dataset[ads/sa.spitzer#0004478976]{NGC~5077}  & E3-4;LINER Sy1.9     &   -5.0 &   2817 &   57.2 & 0.700 &  8.309 &  0.106 & -0.130 &  0.062 &    13.848 &  10.153 \\ 
\dataset[ads/sa.spitzer#0004490496]{NGC~5813}  & E1-2                 &   -5.0 &   1972 &   90.6 & 0.770 &  7.592 &  0.079 & -0.137 & -0.016 &     0.875 &   0.114 \\ 
\dataset[ads/sa.spitzer#0004484352]{NGC~5363}  & E/S0pec[I0?]         &   -3.5 &   1139 &  102.5 & 0.740 &  6.991 &  0.084 & -0.126 &  0.521 &     4.745 &   1.879 \\ 
\dataset[ads/sa.spitzer#0004432640]{NGC~1023}  & SB(rs)0-             &   -3.0 &    637 &  181.0 & 0.380 &  6.318 & -0.036 & -0.143 & -0.007 &     2.708 &   1.833 \\ 
\dataset[ads/sa.spitzer#0004457216]{NGC~4203}  & SAB0-: LINER         &   -3.0 &   1086 &   77.3 & 0.940 &  7.517 &  0.124 & -0.101 &  0.250 &     1.434 &   0.485 \\ 
\dataset[ads/sa.spitzer#0004486656]{NGC~5548}  & (R')SA(s)0/a Sy1.5   &    0.0 &   5149 &   26.3 & 0.940 &  9.547 &  0.204 &  0.546 &  2.545 &     6.128 &   3.619 \\ 
\dataset[ads/sa.spitzer#0004492032]{NGC~6340}  & SA(s)0/a LINER       &    0.0 &   1198 &   51.4 & 0.940 &  8.506 &  0.076 & -0.114 &  0.331 &     0.692 &   0.101 \\ 
\dataset[ads/sa.spitzer#0006629120]{NGC~3031}  & SA(s)ab;LINER Sy1.8  &    2.0 &    -34 &  487.6 & 0.510 &  3.896 &  0.285 & -0.111 &  1.124 &     0.570 &   0.076 \\ 
\dataset[ads/sa.spitzer#0004489728]{NGC~5746}  & SAB(rs)b? sp         &    3.0 &   1724 &  162.2 & 0.260 &  6.934 & -0.119 & -0.102 &  1.256 &     0.335 &   0.040 \\ 
\dataset[ads/sa.spitzer#0004441856]{NGC~3147}  & SA(rs)bc Sy2         &    4.0 &   2820 &   83.1 & 0.820 &  7.523 &  0.319 & -0.053 &  2.451 &     0.511 &   0.064 \\ 
\dataset[ads/sa.spitzer#0004435712]{NGC~1961}  & SAB(rs)c LINER       &    5.0 &   3934 &   78.6 & 0.670 &  7.952 &  0.224 & -0.077 &  2.545 &     0.405 &   0.041 \\ 
\dataset[ads/sa.spitzer#0004488960]{NGC~5669}  & SAB(rs)cd            &    6.0 &   1385 &   55.9 & 0.380 & 10.360 &  0.549 & -0.033 &  2.664 &     0.341 &   0.189 \\ 
\dataset[ads/sa.spitzer#0004494336]{NGC~6689}  & SBcd                 &    6.0 &    467 &   38.1 & 0.580 & 10.336 & -0.084 & -0.045 &  2.103 &     0.109 &   0.033 \\ 
\dataset[ads/sa.spitzer#0006069760]{NGC~\phn300}  & SA(s)d               &    7.0 &    144 &  199.4 & 0.620 &  7.063 &  0.371 & -0.017 &  2.079 &     0.007 &   0.003 \\ 
\dataset[ads/sa.spitzer#0004488192]{NGC~5668}  & SA(s)d               &    7.0 &   1583 &   16.2 & 1.000 & 11.781 &  1.082 & -0.047 &  2.351 &     0.126 &   0.052 \\ 
\dataset[ads/sa.spitzer#0004480512]{NGC~5204}  & SA(s)m HII           &    9.0 &    201 &   42.1 & 0.680 & 10.117 &  0.159 & -0.015 &  1.840 &     0.026 &   0.020 \\ 
\dataset[ads/sa.spitzer#0004434176]{NGC~1156}  & IB(s)m               &   10.0 &    375 &   70.5 & 0.400 &  9.468 &  0.129 & -0.007 &  2.192 &     0.025 &   0.009 \\ 


\enddata



\end{deluxetable}

\end{document}